\documentclass[12pt,times,authoryear]{article}

\usepackage{amssymb,amsfonts,amssymb,subcaption,amsmath,caption,mathtools,float,hyperref,graphicx}
\usepackage{authblk}
\usepackage[round]{natbib}
\title{Local and Global Topics in Text Modeling \\ of Web Pages Nested in Web Sites}
\author[1]{Jason Wang}
\author[1]{Robert E. Weiss}
\affil[1]{Department of Biostatistics,
	Fielding School of Public Health\\
	University of California Los Angeles}

\begin{document}
\maketitle

\begin{abstract}
	Topic models are popular models for analyzing a collection of text documents. The models assert that documents are distributions over latent topics and latent topics are distributions over words. A nested document collection is where documents are nested inside a higher order structure such as stories in a book, articles in a journal, podcasts within an author, or web pages in a web site. In a single collection of documents, topics are global, that is, shared across all documents. For web pages nested in web sites, topic frequencies will likely vary from web site to web site. Within a web site, topic frequencies will almost certainly vary from web page to web page. A hierarchical prior for topic frequencies models this hierarchical structure and specifies a global topic distribution. Web site topic distributions then vary around the global topic distribution and web page topic distributions vary around the web site topic distribution.
	In a nested collection of web pages, some topics are likely unique to a single web site.
	
	Local topics in a nested collection of web pages are topics unique to one web site. 
	For United States local health department web sites, even brief inspection of the text shows local geographic and news topics specific to each health department that are not present in other web sites. 
	Regular topic models that ignore the nesting structure may identify local topics, but do not label those topics as local nor do they explicitly identify the web site owner of the local topic. For web pages nested inside United States local public health web sites, local topic models explicitly label local topics and identifies the owning web site. This identification can be used to adjust inferences about global topics. In the US public health web site data, topic coverage is defined at the web site level after removing local topic words from pages. Hierarchical local topic models can be used to identify local topics, adjust inferences about if web sites cover particular health topics and can be used to study how well health topics are covered. 
\end{abstract}

\section{Introduction}
\label{section}
Topic models have been used to abstract topical information from collections of text documents such as journal abstracts, tweets, and blogs \citep{griffiths_steyvers04,tmblog,socialmediahealth,applications}. Topic models are hierarchical models that define documents as distributions over latent topics and topics as distributions over words. In topic models, each topic is characterized by a vector of word probabilities and each document is characterized by a vector of topic probabilities. Topic-word distributions and document-topic distributions describe the prevalence of words in a topic and topics in a document, respectively. Topics are generally assumed global or shared across all documents \citep{Blei,atm,ctm,rtm,stm}. However, this may not be the case for a nested document collection, where documents are nested inside a higher structure. Examples of nested document collections include articles nested within newspapers, blog posts nested within authors, and web pages nested within web sites. In a nested document collection, some topics may be unique to a group of documents, and we refer to these topics as local topics.

We collected text from web pages nested inside the web sites of local health departments in the United States. We wish to abstract topics from the text and study if and how health topics are covered across web sites. Each web site contains many web pages. Thus, we have a collection of web pages nested within web sites. These web sites have local words and phrases such as geographical names and places that are common within a web site, but are rarely seen on other web sites. Other local words and phrases can be found in local events and local news. 
The content of local topics, how frequent local topic words occur and where local topics are found on a page vary substantially across web sites and web pages. Thus it is difficult to identify local topics a priori and instead we take a probabilistic approach. 

We propose local topic extensions to topic models to accommodate and identify local topics. Local topics can be extensive on individual web pages and can comprise substantial portions of a web site. We do not wish to consider local topics in our desired inferences and so explicitly identifying local topics makes our desired inferences more appropriate. Effectively, local topics are removed from web pages before we make further inferences. We apply our extensions to latent Dirichlet allocation (LDA) models which place Dirichlet priors on topic-word and document-topic distributions \citep{Blei}. 
In a collection of documents, an asymmetric prior on document-topic distributions has been recommended for improved performance over symmetric priors, although symmetric priors remain common and default in applications \citep{priorsmatter,topicmodels}. We expect that a hierarchical asymmetric prior would then fit better for a nested collection of documents.

We consider four models indexed by the number of global topics and apply them to web pages as documents. 
The first model is traditional LDA with an asymmetric prior on document-topic distributions. The second model places a hierarchical asymmetric (HA-LDA) prior on document-topic distributions of the web pages. An asymmetric prior on document-topic distributions accommodates the belief that some topics are more common than others across all web pages and web sites. A hierarchical asymmetric prior further adds that which topics are more common varies from web site to web site. The hierarchical asymmetric prior lets us model the variability of document-topic distributions between web sites. Additionally, the hierarchical asymmetric prior treats web pages as nested inside web sites. 
Our third (LT-LDA) and fourth models (HALT-LDA) introduce local topics, one unique local topic per web site, into the LDA and HA-LDA models. All four models have a fixed maximum number $K$ of global topics. We consider a wide range of values for $K$. 

Nesting in document collections and local topics have been studied in different data settings \citep{rtm,atm,p1,p5,p7,localglobal}. We discuss the similarities and differences in the context of web pages nested in web sites.
Nested document collections can be thought of as a special case of document networks where links are known; web pages of the same web site are linked and web pages of different web sites are not linked \citep{rtm,drtm,srtm}. 
Another type of nesting involves secondary documents nested within a primary document \citep{p1}, such as comments nested within a blog post, and we consider this a separate structure. 
Nested document collections can also be thought of as a collection of different document collections \citep{localglobal} where each web site is itself a document collection.
Relational topic models model the links between any two web pages and are used to predict links to a newly published web page\citep{rtm,drtm,srtm}. We do not need to model links between web pages.
Some models for nested document collections address nesting by modeling multiple levels of document-topic distributions, but do not explicitly model local topics and their topic-word distributions \citep{p5}.
Under the author model in \citet{atm}, local topics are explicitly modeled; however, global topics are not modeled. Under the author-topic model in \citet{atm}, global topics are modeled; however, each web page of a given web site shares the same topic distributions and local topics are not modeled. 
For a single web site or a single document collection, the special words topic model with background distribution (SWB) models a global set of topics, one common web site topic, and a single web page local topic for each web page \citep{p7}. 
The common and distinctive topic model (CDTM) extends SWB and removes web page specific local topics to model multiple web sites or multiple document collections \citep{localglobal}. CDTM models a global set of topics and a separate set of web site local topics for each web site rather than a single web site local topic for each web site. 
We are interested in modeling local topics as a nuisance parameter to adjust our inference; thus, we model a single local topic for each web site to simplify our model and avoid searching for an optimal number of local topics.
Our models additionally place a more flexible asymmetric or hierarchical asymmetric prior on web page topic distributions. 

We show that local topics are not useful for describing words on web pages outside the corresponding local web site. We show this by matching local topics in our HALT-LDA model to global topics in the HA-LDA model and then showing that those matched topics from HA-LDA are not truly global topics but essentially only occur in one web site in the HA-LDA models.

The health department web site data requires additional unique inferences that are not the traditional inferences one would consider when using LDA to analyze a set of reports, newspaper articles, television show transcripts, or books. For the health department web site data we are interested in topic coverage, whether a web site covers a particular topic such as sexually transmitted diseases, emergency preparedness, food safety or heart disease. We are interested in the fraction of web sites that cover a particular topic, and whether a topic is universally covered or not.

Topic coverage has been used to describe the global prevalence of a topic \citep{coverage} or the prevalence of a topic in a document \citep{coverage2}.
However, we are interested in how a web site covers a topic. A health web site contains many web pages that cover different topics, where it dedicates one or a few web pages to a given health topic rather than discusses all health topics across all web pages.
Thus, a topic is covered by a web site if a single page covers the topic and we do not consider a topic covered if many pages have relatively few words from that topic. 
We define topic coverage at the web site level as whether a web site has a page dedicated to that topic, which happens if many or most of the words on a single page are from that topic. 
Further, local topics may be extensive or may be light on various web pages and an extensive local topic coverage should not be allowed to influence a measure of topic coverage at the page level. Thus using models with explicitly identified local topics, we are able to remove words corresponding to the local topic from a page before calculating its coverage. An appropriate topic coverage measure at the web page level needs to calculate fraction of coverage of a particular topic ignoring local topics. Web site coverage should not average across pages, rather web site coverage should consider the supremum of coverage across pages.

Section \ref{topicmodels} defines notation and our four models. Section \ref{modelinference} discusses computation and  inference. Section \ref{healthdata} introduces our motivating data set in greater detail and section \ref{results} lays out our analysis and illustrates the conclusions that are of interest for this data and the conclusions that local models allow for. The paper closes with discussion.

\section{Topic Models for Nested Web pages}
\label{topicmodels}
In a collection of web sites, we define a document to be a single web page. Thus, we refer to the document-topic distribution of a web page as the web page-topic distribution. Web sites are indexed by $i = 1,\ldots, M$ and web pages nested within web sites are indexed by $j = 1,\ldots, M_i$.
Words $w_{ijh}$ on a page are indexed by $h = 1,\ldots, N_{ij}$ and the set of unique words across all web sites and web pages are indexed by $v = 1, \ldots, V$ where $V$ is the number of unique words or the size of the vocabulary. The number of global topics $K$, indexed by $k = 1,\ldots,K$, is assumed fixed and known prior to modeling as in latent Dirichlet allocation. Table \ref{tab:desc} details notation used in our models.

\subsection{Latent Dirichlet Allocation}
\label{lda}
Latent Dirichlet allocation (LDA) asserts that topics are global and their topic-word distributions are drawn from a Dirichlet prior.
For Dirichlet distributed parameters $\phi_k$ we use the parameterization
\begin{align*}
	\phi_k \sim \text{Dirichlet}(c_{\beta}\beta),
\end{align*}
where $\phi_k$ is a $V$-vector of probabilities $\phi_{k,v}$ such that $\sum_{v=1}^{V} \phi_{kv} = 1$, $0\leq\phi_{kv}\leq1$, $c_{\beta} > 0$ is a scale parameter, and $\beta$ is a $V$-vector of parameters $\beta_v$ such that a priori $E[\phi_{k}|c_{\beta}\beta] = \beta$, $\sum_{v=1}^{V}\beta_{v} = 1$, and $0\leq\beta_{v}\leq1$. Each web page $j$ in web site $i$ has web page-topic distribution denoted by a $K$ vector of probabilities $\theta_{ij}$ with a Dirichlet($c_\alpha\alpha$) prior. Topic $k$ has a topic-word multinomial distribution parameterized by a $V$-vector of probabilities $\phi_k$ a priori distributed as Dirichlet($c_{\beta}\beta$). Words have a latent topic $z_{ijh}$. The LDA model is 
\begin{align*}
	\theta_{ij}|c_{\alpha}\alpha & \sim \text{Dirichlet}(c_{\alpha}\alpha), \\
	\phi_k|c_{\beta}\beta & \sim \text{Dirichlet}(c_{\beta}\beta), \\
	z_{ijh}|\theta_{ij} & \sim \text{Categorical}(\theta_{ij}), \\
	w_{ijh}|\phi_{z_{ijh}} & \sim \text{Categorical}(\phi_{z_{ijh}}).
\end{align*}
Documents in LDA are characterized by a distribution over all $K$ topics, thus, LDA has $K$ global topics and no local topics.
\begin{table*}
	\caption{Model notation with definitions.}
	\label{tab:desc}
	\centering
	\begin{tabular}{rl}
		\hline
		Notation & Description \\
		\hline
		$i$ & Web site index, $i=1,\ldots,M$ \\
		$j$ & Web page index, $j = 1,\ldots,M_i$ \\
		$h$ & Word index, $h = 1,\ldots,N_{ij}$ \\
		$M$ & Number of web sites \\
		$M_i$ & Number of pages in web site $i$ \\
		$N_{ij}$ & Number of words in page $j$ in web site $i$ \\
		$K$ & Number of global topics \\
		$L$ & Number of local topics \\
		$L_{i}$ & Number of local topics in web site $i$ \\
		$V$ & Number of unique words in the vocabulary \\
		$\theta_{ij}$ & Page-topic distribution of web site $i$ web page $j$\\
		$\psi_{i}$ & Local topic-word distribution of web site $i$ \\
		$\phi_{k}$ & Global topic-word distribution of topic $k$ \\
		$w_{ijh}$ & Word $h$ of page $j$ in web site $i$ \\
		$z_{ijh}$ & Topic choice of the $h$th word of page $j$ in web site $i$ \\
		\hline
	\end{tabular}	
\end{table*}
\subsection{Local Topics}
\label{localtopics}

Now we introduce $L$ local topics distributed among $M$ web sites, such that each web site $i$ contains $L_i$ local topics and $L = \sum_{i=1}^{M}L_{i}$. We let $l = 1,\ldots,L_i$ index local topics in web site $i$. The web page-topic distribution, $\theta_{ij}$, for page $j$ in web site $i$ is now a $(K + L_i)$-vector of probabilities. The topic-word distribution $\psi_{il}$ for each local topic is still a $V$ vector of probabilities with a Dirichlet($c_{\gamma}\gamma$) prior. We define the $(K+L_i)\times V$ array, $\Phi_{i} = \{\phi_1, \ldots, \phi_K,\psi_{i1},\ldots,\psi_{iL_i}\}$, as the combined set of global and local topic-word distributions for web site $i$.
The LT-LDA model is then 
\begin{align*}
	\theta_{ij}|c_{\alpha}\alpha &\sim \text{Dirichlet}(c_{\alpha}\alpha), \\
	\psi_{il}|c_{\gamma}\gamma &\sim \text{Dirichlet}(c_{\gamma}\gamma), \\ 
	\phi_k|c_{\beta}\beta &\sim \text{Dirichlet}(c_{\beta}\beta), \\
	z_{ijh}|\theta_{ij} &\sim \text{Categorical}(\theta_{ij}), \\
	w_{ijh}|\Phi_{iz_{ijh}} &\sim \text{Categorical}(\Phi_{iz_{ijh}}).
\end{align*}
The shared prior parameter $\alpha$ requires that $L_1 = \ldots = L_M$; however, this can be generalized so that each web site $i$ has a separate and appropriate prior for $\theta_{ij}$. In our applications with local topics, we choose $L_i = 1$ for all $i = 1, \ldots, M$ assuming that most web sites have one local topic that places high probability on geographical names and places.

\subsection{Hierarchical Asymmetric Prior}
\label{hierarchicalap}

A symmetric prior Dirichlet($c_{\alpha}\alpha$) for web page-topic distributions $\theta_{ij}$ is such that $c_{\alpha}\alpha = d\times\{1,\ldots,1\}$ for some constant $d$ and describes a prior belief about the sparsity or spread of page-topic distributions. A smaller $d$ describes the prior belief that web pages have high probability for a small number of topics and low probability for the rest, while a larger $d$ describes the prior belief that web pages have more nearly equal probability for all topics. A single asymmetric prior Dirichlet($c_{\alpha}\alpha$), such that $c_{\alpha}\alpha = \{d_1,\ldots,d_{K+1}\}$ where not all $d_k$ are equal, accommodates the belief that topics or groups of words with larger $d_k$ will occur more frequently across all pages than topics with smaller $d_k$.

For a nested document collection, we extend the belief that different topics occur more frequently to multiple levels. Thus a given topic will have different probabilities in different web sites, and also, that topic's probability will vary across web pages within a web site.
Globally, some topics are more common than others and while we start with a symmetric Dirichlet prior for the unknown global-topic distribution, the global-topic distribution will be asymmetric. Locally, each web site has its own set of common and uncommon topics with the web site-topic distribution centered at the global-topic distribution. Finally each web page within a web site will have their own common and uncommon topics and web page-topic distribution are centered around the web site-topic distribution. We extend the LDA model in section \ref{lda} by placing a hierarchical asymmetric prior on web page-topic proportions such that web pages nested within web sites share commonalities. We first place a $\text{Dirichlet}(c_{\alpha}\alpha_i)$ prior on web page-topic distribution $\theta_{ij}$, such that each web site has a $(K+1)$-vector of parameters $\alpha_i$ so that a priori $E[\theta_{ij}|c_{\alpha}\alpha_i] = \alpha_i$. We next place a $\text{Dirichlet}(c_{0}\alpha_0)$ prior on web site-topic distributions $\alpha_i$. The HA-LDA model is 
\begin{align*}
	\theta_{ij}|c_{\alpha}\alpha_i &\sim \text{Dirichlet}(c_{\alpha}\alpha_i), \\
	\alpha_i|c_{0}\alpha_0 &\sim \text{Dirichlet}(c_{0}\alpha_0),\\
	\phi_k|c_{\beta}\beta &\sim \text{Dirichlet}(c_{\beta}\beta),\\
	z_{ijh}|\theta_{ij} &\sim \text{Categorical}(\theta_{ij}),\\
	w_{ijh}|\phi_{z_{ijh}} &\sim \text{Categorical}(\phi_{z_{ijh}}).
\end{align*}
We further place Gamma priors on $c_{\alpha}$ and each element of $c_0\alpha_{0,k}$. 
Combining the hierarchical asymmetric prior with local topics, the HALT-LDA model is
\begin{align*}
	\theta_{ij}|c_{\alpha}\alpha_i &\sim \text{Dirichlet}(c_{\alpha}\alpha_i), \\
	\alpha_i|c_{0}\alpha_0 &\sim \text{Dirichlet}(c_{0}\alpha_0),\\
	\psi_{il}|c_{\gamma}\gamma &\sim \text{Dirichlet}(c_{\gamma}\gamma), \\ 
	\phi_k|c_{\beta}\beta &\sim \text{Dirichlet}(c_{\beta}\beta),\\
	z_{ijh}|\theta_{ij} &\sim \text{Categorical}(\theta_{ij}),\\
	w_{ijh}|\Phi_{iz_{ijh}} &\sim \text{Categorical}(\Phi_{iz_{ijh}}).
\end{align*}

\subsection{Prior Parameter Specification}
\label{priorparameters}
We place an asymmetric prior on $\alpha$ and a Gamma prior on $c_{\alpha}$ in LDA and LT-LDA. Therefore the difference between LDA and LT-LDA is the addition of local topics and the difference between LDA and HA-LDA is the use of a hierarchical asymmetric prior over a single asymmetric prior. We compare these models to study the impact of each extension. We also compare these models to a model with both a hierarchical asymmetric prior and local topics (HALT-LDA). We specify prior parameters to accommodate sparse mixtures of topics. In LDA and LT-LDA, we place priors
\begin{align*}
	c_{\alpha} &\sim \text{Gamma}(a_\alpha,b_\alpha), ~~~~ a_\alpha = b_\alpha = 1,
	\\
	\alpha &\sim \text{Dirichlet}(\{1/K^*,\ldots,1/K^*\}),
\end{align*}
where we use the shape-rate parameterization of the Gamma distribution with mean $a_\alpha b_\alpha$ and where $K^*=K$ in LDA and $K^*=K+1$ in LT-LDA.
In HA-LDA and HALT-LDA, we treat $c_0\alpha_{0,k}$ as a single parameter and place priors
\begin{align*}
	c_{\alpha} &\sim \text{Gamma}(a_\alpha,b_\alpha), ~~~~ a_\alpha = b_\alpha = 1,
	\\
	c_0\alpha_{0,k} &\sim \text{Gamma}(1,1).
\end{align*}
We generated 100,000 sets of $c_0\alpha_{0,k}$ for $K = 50$.  This generates a 
largest order statistic for $\alpha_{i,k}$ of 0.09 with a standard deviation of 0.04. At K = 100, the largest order statistic is 0.05 with a standard deviation of 0.02. 
The largest order statistic from the prior differs from the overall local topic prevalence in our results in section \ref{applications}; however, a priori, this result for the highest order statistic was reasonable. Later order statistics were reasonably modeled with Gamma(1,1)
We expect each topic to place high probability above 0.02 on a small subset of words but do not expect any words to have high probability across all topics. Therefore, we place a symmetric prior over topic word distributions, $\phi_k$ and $\psi_i$. 
The priors are fixed such that $c_{\beta}\beta = c_{\gamma}\gamma = \{0.05,\ldots,0.05\}$. Sensitivity analysis in section A.2 of the web appendix shows that conclusions from  HALT-LDA are robust to deviations from our choice of $c_\beta$, $c_\gamma$, and $a_\alpha$. 

\section{Computation and Inference for Hierarchical Topic Models}
\label{modelinference}
The general goal of inference in hierarchical topic models is to estimate the topic-word distributions, $\phi_k$ and $\psi_{i}$, and web page-topic distributions, $\theta_{ij}$. We use Markov chain Monte Carlo (MCMC) to sample from the posterior, where unknown parameters are sequentially sampled conditional on current values of all other unknown parameters. We outline the sampler for the most complex model, HALT-LDA, where each web site has $L_i = 1$ local topic $\psi_i$. 

Let $W$ and $Z$ be ragged arrays of identical structure, with one element $w_{ijh}$ and $z_{ijh}$ for every word $h$ in web page $j$ from web site $i$. The $ijh$ element of $W$ corresponding to the $ijh$ word identifies the index from $1$ to $V$ of that word, and the corresponding element $Z_{ijh}$ of $Z$ identifies the topic assigned to that word. As $Z$ is latent, it is sampled and will change at every iteration of the MCMC algorithm. Let $\alpha$ be the set of all web site-topic distributions $\alpha_i$ and similarly, let $\theta$, $\phi$, and $\psi$ be the sets of all $\theta_{ij}$, $\phi_k$, and $\psi_i$. Then the joint prior density of all unknown parameters and data is 
\begin{align*}
	\MoveEqLeft P(W,Z,\phi,\psi,\theta,c_{\alpha},\alpha,c_0\alpha_0) = \\
	& P(W|Z,\phi,\psi) P(Z|\theta) P(\theta|c_{\alpha},\alpha) P(\alpha|c_0\alpha_0) P(c_0\alpha_0) P(\phi) P(\psi).
\end{align*}
Dirichlet-multinomial conjugacy allows us to algebraically integrate out $\phi_k$, $\psi_{il}$, and $\theta_{ij}$ from the posterior. We are left to sample topics $z_{ijh}$ of each word $w_{ijh}$, scale parameter $c_{\alpha}$, and web site-topic distributions $\alpha_i$ and their prior parameters $c_0\alpha_{0,k}$.

Let $n_{k,v}$, $p_{i,v}$, and $m_{ij,k}$ be counts that are functions of $Z$ and $W$. These counts vary from iteration to iteration as they depend on $Z$. Let $n_{k,v}$ be the total count of word $v$ assigned to topic $k$, let $p_{i,v}$ be the count of word $v$ from the single local topic of web site $i$, and let $m_{ij,k}$ be the count of words from topic $k$ in page $j$ of web site $i$. Let the superscript $^{-}$ on counts $n_{k,w_{ijh}}^-$, $m_{ij,k}^-$, and $p_{i,w_{ijh}}^-$ indicate that the counts exclude word $w_{ijh}$. Similarly, let $Z^{-}$ be the set of topic indices $Z$ excluding word $w_{ijh}$. Then the sampling density for $z_{ijh}$ conditioned on scale parameter $c_{\alpha}$, web site-topic distribution $\alpha_i$, and the remaining topics indices $Z^{-}$
is 
\begin{align*}
	\MoveEqLeft P(z_{ijh} = k|Z^{-},c_{\alpha},\alpha_i,w_{ijh}) \propto \\
	& (m_{ij,k}^- + c_{\alpha}\alpha_{i,k}) \times
	\bigg(\frac{n_{k,w_{ijh}}^- + \beta_v}{\sum_{v=1}^{V}n_{k,v}^- + \beta_v} \bigg)^{1_{k\leq K}}\times
	\bigg(\frac{p_{i,w_{ijh}}^- + \gamma_v}{\sum_{v=1}^{V}p_{i,v}^- + \gamma_v} \bigg)^{1_{k=K+1}},
\end{align*}
where $1_{k\leq K}$ is an indicator function that is one if $k$ is a global topic and zero if $k$ is a local topic and $1_{k=K+1} = 1- 1_{k\leq K}$. To sample web site-topic distribution $\alpha_i$ we use a data augmentation step with auxiliary variables $\lambda_{ij,k}$ with conditional density 
\begin{align*}
	P(\lambda_{ij,k}|Z,c_{\alpha}\alpha_{i,k},\lambda_{-(ij,k)}) = \frac{\Gamma(c_{\alpha}\alpha_{i,k})}{\Gamma(c_{\alpha}\alpha_{i,k}+m_{ij,k})} |s(m_{ij,k},\lambda_{ij,k})| (c_{\alpha}\alpha_{i,k})^{\lambda_{ij,k}},
\end{align*}
where $s(\cdot,\cdot)$ is the Stirling number of the first kind.
This step allows posterior draws of web site-topic distribution $\alpha_i$ from a Dirichlet($c_{0}\alpha_0 + \sum_{j=1}^{M_i}\lambda_{ij,k}$) \citep{hdp}. Parameters $c_{\alpha}$ and $c_0\alpha_{0,k}$ are sampled using Metropolis-Hastings.

We estimate conditional means of the multinomial parameters $\phi_k$, $\psi_{i}$, and $\theta_{ij}$ for each MCMC sample, as is common in using MCMC sampling in topic models. Let superscript $(q)$ indicate a count, estimate, or sample from iteration $q$ of the MCMC sample.  Each iteration $q$ samples a topic index for every word. The conditional estimate of the global topic-word proportions $\phi_k$ at iteration $q$ is given by the conditional posterior mean
\begin{align*}
	\bar{\phi}_{k,v}^{(q)} &= \frac{c_{\beta}\beta_v + n_{k,v}^{(q)}}{\sum_{v=1}^{V} c_{\beta}\beta_v + n_{k,v}^{(q)}}.
\end{align*}
Similarly, the conditional posterior means for the local topic-word mixture $\psi_{i,v}$ and web page-topic mixtures $\theta_{ij,k}$ at iteration $q$ are
\begin{align*}
	\bar{\psi}_{i,v}^{(q)} &= \frac{c_{\gamma}\gamma_v + p_{i,v}^{(q)}}{\sum_{v=1}^{V} c_{\gamma}\gamma_v + p_{i,v}^{(q)}},
	\\
	\bar{\theta}_{ij,k}^{(q)} &= \frac{c_{\alpha}\alpha_{ik} + m_{ij,k}^{(q)}}{\sum_{k=1}^{K+1} c_{\alpha}\alpha_{ik} + m_{ij,k}^{(q)}}.
\end{align*}	

We perform a 10-fold cross validation to compare fits of LDA, LT-LDA, HA-LDA, and HALT-LDA to the health departments web site data. Each fold splits the data randomly, holding out 20\% of the pages from a web site and using the other 80\% of pages for MCMC sampling. 
For each sample $q$ we calculate and save
conditional posterior means $\bar{\phi}_{k,v}^{(q)}$ and $\bar{\psi}_{i,v}^{(q)}$ and save the sampled $c_\alpha$ and $\alpha_i^{(q)}$.
We save results from 500 MCMC iterations after a burn-in of 1500.
We calculate an estimate for scale parameter $c_\alpha$ and probability vector $\alpha_i$ by averaging over the 500 saved samples.
We calculate an estimate for topic-word probabilities $\phi_{k,v}$ and $\psi_{i,v}$ by averaging over 500 conditional posterior means.
We use the estimates to calculate the held-out log likelihood of held-out pages given $c_\alpha$, $\alpha_i$,  $\phi_{k,v}$, and $\psi_{i,v}$.
We use the left-to-right particle filtering algorithm for LDA to approximate held-out log likelihoods \citep{Wallach2009}. 
Wallach's left-to-right algorithm sequentially samples topic indices and calculates log likelihood components of each word from left to right. The algorithm decomposes the probability of a held-out word to a sum over joint probabilities of a held-out word and topic indices of previous words in the same document. The algorithm has been described by \citet{particlegibbs} as a particle-Gibbs method. We provide a brief summary of the algorithm applied to HALT-LDA in section A.3 of the web appendix.
Held-out log likelihoods are averaged over the cross-validation sets and used to identify a reasonable choice for the number of global topics $K$ and to compare between the LDA, LT-LDA, HA-LDA, and HALT-LDA. We analyze a final HALT-LDA model with 1,000 samples after a burn-in of 1,500 samples.

\section{Health Department Web Site Data}
\label{healthdata}
The National Association of County and City Health Officials maintains a directory of local health departments (LHD) in the United States that includes a URL for each department web site \citep{naccho}. We scrape each web site for its textual content using Python and Scrapy \citep{python,scrapy}. All web sites were scraped during November of 2019. We remove text items that occur on nearly every page, such as titles or navigation menus. Pages with fewer than 10 words are removed. Common English stop words, such as `the', `and', `them', and non-alphabet characters are removed, and words are \textit{stemmed}, e.g.\ `coughing' and `coughs' are reduced to `cough'. Uncommon words, which we define as words occurring in fewer than 10 pages across all web sites, are removed. Due to computation time of MCMC sampling, a subset of 20 web sites with fewer than 100 pages each were randomly selected to use in our analyses. The dataset analyzed had 124,491 total words with $V = 1614$ unique words across 923 pages. At $K = 60$ it takes approximately 65 minutes to run 1000 total iterations with HALT-LDA with an Intel Core i7-6700 processor.

\section{Results}
\label{results}
\begin{figure}
	\centering
	\includegraphics[scale = .65]{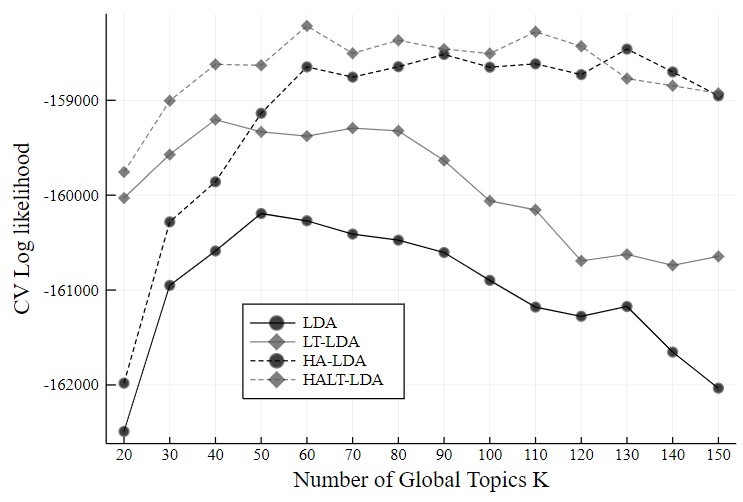}
	\caption{Plot of 10-fold cross validated (CV) held-out log likelihood by different number of global topics $K$. }
	\label{fig:likelihoodcomparison20}
\end{figure}

The 10-fold cross validated held-out log likelihoods are plotted against the number of global topics $K$ in Figure \ref{fig:likelihoodcomparison20} for the four models: LDA, LT-LDA, HA-LDA, and HALT-LDA. For every fixed number of global topics $K$, our extensions LT-LDA, HA-LDA and HALT-LDA outperform LDA. 
At smaller $K$, because they also include 20 local topics, HALT-LDA and LT-LDA allow more total topics compared to HA-LDA and LDA. Thus, we expect and see that models with local topics perform better at a smaller number of global topics $K$.
The consistent improvement in log likelihood from LDA to LT-LDA indicates that local topics exist and that web pages in a web site do indeed share a local topic.
However, the improvement from HA-LDA to HALT-LDA decreases as $K$ increases.
This is because the nested asymmetric prior is a flexible prior that can accommodate local topics though it does not formally identify specific topics as local. It allows pages of a web site to share commonalities, such as high probability in its local topic and low probability in local topics of other web sites. The HALT-LDA cross-validated log likelihoods peak slightly higher and at smaller $K$, while HA-LDA peaks at larger $K$. Both these models support a larger number of topics than their counterparts without a hierarchical asymmetric prior. 
The results suggests that LT-LDA, HA-LDA, and HALT-LDA model web pages nested in web site better than LDA, and local topics allow us to specify a smaller number of global topics with similar or better performance.
In later inference for the public health departments, we are not interested in the local topics except to remove words corresponding to local topic from pages before further calculations. Therefore, it is much more useful to use the LT models which automatically identify local topics to more easily make inferences only about global topics.

\subsection{Matching and Comparing Local Topics}
\label{matching}
We match local topics in HALT-LDA with $K = 60$ to global topics in HA-LDA with $K=90$ to illustrate the existence of local topics and their high prevalence within a single web site relative to their prevalence in other web sites. 
We choose $K=60$ for HALT-LDA where log likelihood peaks and choose $K = 90$ where HA-LDA performs nearly at its peak at $K = 130$ but is closer to HALT-LDA in total number of topics.
We compare two methods for matching topics; a rank based method and a probability based method. The rank based method finds topics in HA-LDA that have similar sets of word ranks as a local topic in HALT-LDA while the probability based method finds topics in HA-LDA that have similar word probabilities as a local topic in HALT-LDA.
Let $R^{\text{(HA)}}_{k,v}$ denote the rank of word $v$ in topic $k$ from HA-LDA and let $R^{\text{(HALT)}}_{i,v}$ denote the rank of word $v$ in local topic $i$ from HALT-LDA. For the rank based method, the matched topic index in HA-LDA for local topic $i$ is
\begin{equation}
	\underset{k}{\arg\min}\sum_{v=1}^{V}|R_{i,v}^{(HALT)} - R_{k,v}^{(HA)}| \label{match1}.
\end{equation}
Define $\psi_{i,v}^{(HALT)}$ as the local topic-word probability for web site $i$ and word $v$ in HALT-LDA and define $\phi_{k,v}$ as the topic-word probability for topic $k$ and word $v$ in HA-LDA.
By the probability based method, the matched topic index in HA-LDA for local topic $i$ is
\begin{equation}
	\underset{k}{\arg\min}\sum_{v=1}^{V}(\psi_{i,v}^{(HALT)} - \phi_{k,v}^{(HA)})^2 \label{match2}.
\end{equation}

Topics generally place higher probability on a small subset of words while placing small probability on the majority of words.
We may want to consider only the most probable subset of words in our calculations in equation \ref{match1} and equation \ref{match2} if we define topics by their most probable words.
Thus, we consider limiting the summations to the subset of most common words. Define $T^{(10)}_i$ as the indices of the top 10 words from local topic $i$ in HALT-LDA. Then the calculations for rank based and probability based matching are respectively
\begin{align*}
	\underset{k}{\arg\min}&\sum_{v\in T^{(10)}_i}|R_{i,v}^{(HALT)} - R_{k,v}^{(HA)}|,
	\\
	\underset{k}{\arg\min}&\sum_{v\in T^{(10)}_i}(\psi_{i,v}^{(HALT)} - \phi_{k,v}^{(HA)})^2.
\end{align*}
We estimate topic-word probabilities by averaging across 1,000 conditional posterior means and match using those estimates.
For each web site $i$, we matched one topic in HA-LDA to local topic $i$ in HALT-LDA. Thus, there are 20 matched local topics in HA-LDA, one for each web site. For a given web site, we refer to the matched local topic that belongs to the web site as the \textit{correct local} topic and the remaining 19 matched local topics as \textit{other local} topics. 

\begin{figure}
	\centering
	\includegraphics[scale = .65]{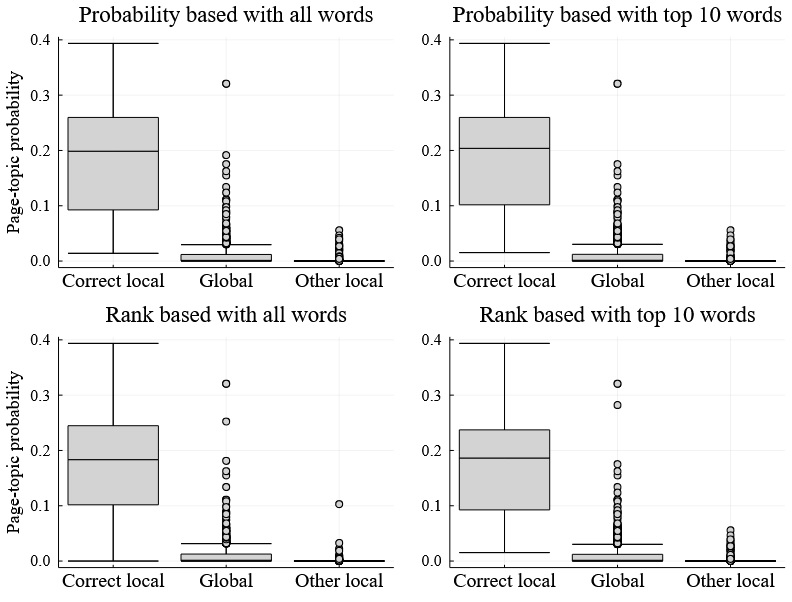}
	\caption{Boxplots of the web site average web page-topic distributions $\bar{\theta}_{i\cdot,k} = \frac{1}{M_i}\sum_{j=1}^{M_i}\theta_{ij,k}$ of global topics and  matched local topics in HA-LDA. `Correct local' shows the distribution of $\bar{\theta}_{i\cdot,k}$, where topic $k$ has been matched to web site $i$'s local topic in HALT-LDA. `Other local' shows the distribution of $\bar{\theta}_{i\cdot,k}$, where topic $k$ is a local topic but not the matched local topic. Global shows the distribution of $\bar{\theta}_{i\cdot,k}$ for the remaining topics $k$.}
	\label{fig:correctother20}
\end{figure}

Web site averages, $\bar{\theta}_{i\cdot,k} = \frac{1}{M_i}\sum_{j=1}^{M_i}\theta_{ij,k}$, of web page-topic distributions are calculated by averaging estimates across pages of a web site. Thus in HA-LDA there are 20 averages that correspond to \textit{correct local} topics, 380 averages that correspond to \textit{other local} topics, and 1400 averages that correspond to the remaining global topics.
Figure \ref{fig:correctother20} plots boxplots of web site average probabilities for \textit{correct local} topics, \textit{other local} topics, and global topics plotted in between as a reference.
The first row shows the probability based methods and the second row shows the rank based methods. The first column are methods using all words and the second column using top 10 words.
There is extreme localization of local topics in HA-LDA regardless of topic matching method. \textit{Correct local} topics typically have high web site average probabilities, global topics have lower averages, and \textit{other local} topics have the lowest averages, with most nearly 0. 

\subsection{Topic Model Output and Applications}
\label{applications}
\begin{figure}
	\includegraphics[scale = .65]{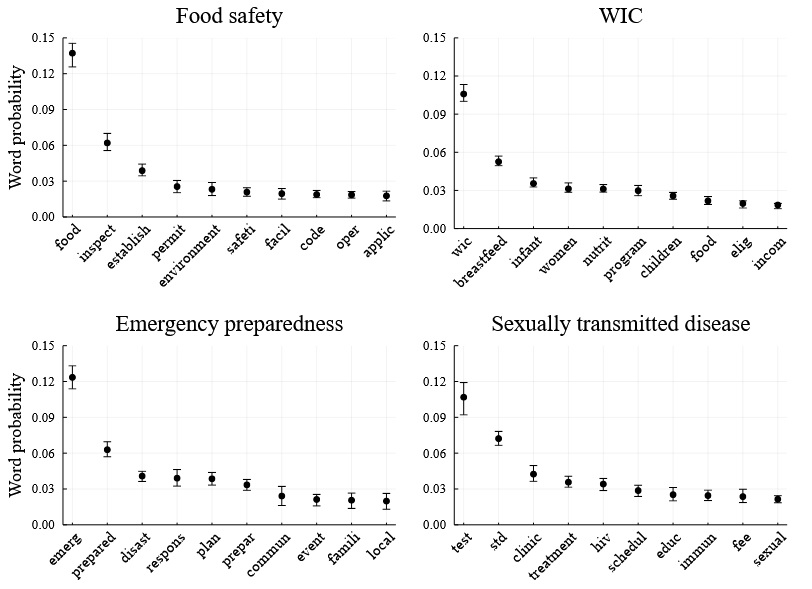}
	\caption{Median and 95\% intervals of conditional posterior means of word probabilities for the ten most probable words in four health topics. }
	\label{fig:health20}
\end{figure}

Table \ref{tab:topics} lists the ten most probable words for the most prevalent global topic and for another 9 health topics from among the top 20 highest probability topics in HALT-LDA for $K=60$. 
We label each topic after inspecting its most probable words. The prevalence column shows the average probability of a topic across all web pages and web sites. The most prevalent (5.4\%) topic has top words \textit{inform, provid, contact, please, requir, call, need, must, click, may} that generally describe getting information and contacting the public health department. The cumulative prevalence of all 60 global topics is 82\%, with 18\% in local topics. Thus, the local topic in each web site generally accounts for a large proportion of text. Four health topics we use in our later analysis are food safety, Special Supplemental Nutrition Program for Women, Infants, and Children (WIC), emergency preparedness, and sexually transmitted disease. Estimates and 95\% intervals of conditional posterior means for word probabilities of these topics' ten most probable words are plotted in Figure \ref{fig:health20}. The word probabilities for the ten most probable words are much larger than the average probability 1/1614. 

\begin{table*}
	\caption{The ten highest probability words for the most common topic (General) and nine health topics from HALT-LDA for $K = 60$. Topic labels in the first column are manually labeled and the prevalence is the average probability across all web pages and web sites. Means and 95\% credible intervals for the probabilities of the words for the 4 health topics in boldface are plotted in Figure \ref{fig:health20}.}
	\label{tab:topics}
	\centering
	\begin{tabular}{ccl}
		\hline
		Label & Prevalence & \multicolumn{1}{c}{Top 10 words}\\
		\hline
		General & 5.4\% & \textit{inform, provid, contact, pleas, requir, }\\
		&& ~~\textit{call, need, must, click, may} \\
		Disease prevention & 3.3\% & \textit{diseas, prevent, risk, caus, use,}\\
		&& ~~\textit{includ, year, effect, peopl, also} \\
		\textbf{Food safety}  & 2.9\% & \textit{food, inspect, establish, permit, environment, }\\
		&& ~~\textit{safeti, facil, code, oper, applic} \\	
		\textbf{WIC}  & 2.7\% & \textit{wic, breastfeed, infant, women, nutrit, }\\
		&& ~~\textit{program, children, food, elig, incom} \\
		Vaccinations  & 2.0\% & \textit{immun, vaccin, adult, children, child,}\\
		&& ~~\textit{schedul, flu, appoint, clinic, diseas} \\
		Breast cancer  & 1.9\% & \textit{test, women, clinic, screen, famili, }\\
		&& ~~\textit{pregnanc, plan, breast, cancer, exam} \\		
		\textbf{Emergency preparedness}  & 1.8\% & \textit{emerg, prepared, disast, respons, plan, }\\
		&& ~~\textit{prepar, commun, event, famili, local} \\
		Hospital Care  & 1.7\% & \textit{care, patient, provid, medic, nurs}\\
		&& ~~\textit{physician, treatment, visit, hospit, includ} \\
		\textbf{Sexually transmitted disease}  & 1.5\% & \textit{test, std, clinic, treatment, hiv, }\\
		&& ~~\textit{schedul, educ, immun, fee, sexual} \\
		Family Program  & 1.4\% & \textit{child, children, famili, parent, program}\\
		&& ~~\textit{visit, home, babi, help, hand} \\	
		\hline
	\end{tabular}	
\end{table*}

Table \ref{tab:localtopics} lists the five most probable words for each of the $M = 20$ local topics. Most local topics contain a geographical name or word among its top five words. The local topic in web site 7 has top words related to food sanitation inspection because web site 7 contains 14 pages dedicated to reports for monthly inspections and another 16 pages related to food protection and food sanitation out of a total of 86 pages. 
The local topic in web site 13 has top words related to food sanitation inspection because 11 of its 30 pages mention food inspections.
In Table \ref{tab:topics}, food safety is a global topic that shares similar words. We further investigate the food safety topic later in our analysis.
Web site 9 is the only web site with several pages containing placeholder text, i.e.\ lorem ipsum or nonsensical Latin, which account for the top words in its local topic. Web site 15 has two large pages each with about 3000 words describing job openings which account for the top words in its local topic. Other than the local topic in web site 7 and 13, no other local topic is similar to the global topics in Table \ref{tab:topics}.
\begin{table*}
	\caption{Top five highest probability words in local topics from HALT-LDA for $K=60$. Most local topics include a geographical name or word among the top five words. }
	\label{tab:localtopics}
	\centering
	\begin{tabular}{rlll}
		\hline
		& Location & State & Top 5 Words (local topic)\\
		\hline
		1& Elkhorn Logan Valley & Nebraska & \textit{month, nation, awar, elvphd, day}\\	
		2& Sandusky County & Ohio & \textit{sanduski, ohio, fremont, street, read}\\	
		3& Ford County & Illinois & \textit{ford, program, illinoi, bird, press}\\	
		4& Loup Basin & Nebraska & \textit{loupbasin, loup, basin, nebraska, program}\\	
		5& Wayne County & Missouri & \textit{center, wayn, creat, homestead, back}\\	
		6& Greene County & Iowa & \textit{green, medic, center, care, therapi}\\	
		7& Bell County & Texas & \textit{report, inspect, food, retail, octob}\\	
		8& Moniteau County & Missouri & \textit{moniteau, missouri, center, requir, map}\\	
		9& Williams County & Ohio & \textit{phasellu, sed, dolor, fusc, odio}\\
		10& Harrison and Clarksburg & West Virginia & \textit{alert, harrison, clarksburg, subscrib, archiv}\\
		11& Oldham County & Kentucky & \textit{oldham, kentucki, click, local, resourc}\\
		12& Boyle County & Kentucky & \textit{boyl, bag, item, bed, home}\\
		13& Dallas County & Missouri & \textit{buffalo, routin, dalla, food, inspect}\\
		14& Shelby County & Tennessee & \textit{sschd, ohio, shelbycountyhealthdeptorg,}\\ 
		&&& \textit{email, shelbi}\\
		15& Taney County & Missouri & \textit{averag, normal, assur, commun, exposur}\\
		16& Monroe County & Missouri & \textit{monro, phone, email, map, fax}\\
		17& Three Rivers District & Kentucky & \textit{river, three, district, kentucki, local}\\
		18& Central District & Nebraska & \textit{central, district, permit, resourc, island}\\
		19& Levy County & Florida & \textit{florida, updat, weekli, month, april}\\
		20& Ozark County & Missouri & \textit{ozark, contact, info, home, box}\\
		\hline
	\end{tabular}	
\end{table*}

Web sites 7, 9, and 15 have global topics that appear to be local topics for these web sites. The global topic with top words \textit{taney, report, commun, anim, outreach} may be a second local topic for web site 15 as it is related to a common news block in several web pages. Similarly the global topic with top words \textit{william, ohio, dept, divis, inform} and the global topic with top words \textit{nbsp, bell, district, texa, director} may be second local topics for web sites 9 and 7. These three global topics were less prevalent within the respective web sites than the local topics discovered by the model.
Additionally, we found two other global topics with top words \textit{green, center, medic, foundat, jefferson} and \textit{shall, section, ordin, dalla, person} that may be second local topics for web site 6 and 13. 
The global topic with top words \textit{green, center, medic, foundat, jefferson} has nearly the prevalence within web site 6 as the local topic of web site 6.
The global topic with top words \textit{shall, section, ordin, dalla, person} is more prevalent in web site 13 than the local topic of web site 13. However, the identified local topic with top words \textit{buffalo, routin, dalla, food, inspect} has more local words specific to web site 13 than the global topic. Our model either identifies the most prevalent local topic or the local topic with more local words.

We model public health web sites using topic models to understand how local health departments cover health topics online. In a web site, multiple health topics may be covered and it is more reasonable to dedicate a single or handful of web pages to a given health topic rather than have every web page discuss all health topics. Rather than comparing web site average probabilities of a given topic, we compare topic coverage. Informally, topic coverage measures whether a web site has at least one dedicated page on a given topic. Formally, we define coverage of topic $k$ in web site $i$ as the largest web page-topic probability $\theta_{ij,k}$ across all $j = 1, \ldots, M_i$ pages,
\begin{align*}
	\underset{j}{\max} ~ \theta_{ij,k}.
\end{align*}
We use topic coverage to help identify common health topics that may be missing in a web site.

\begin{figure}
	\includegraphics[scale = .6]{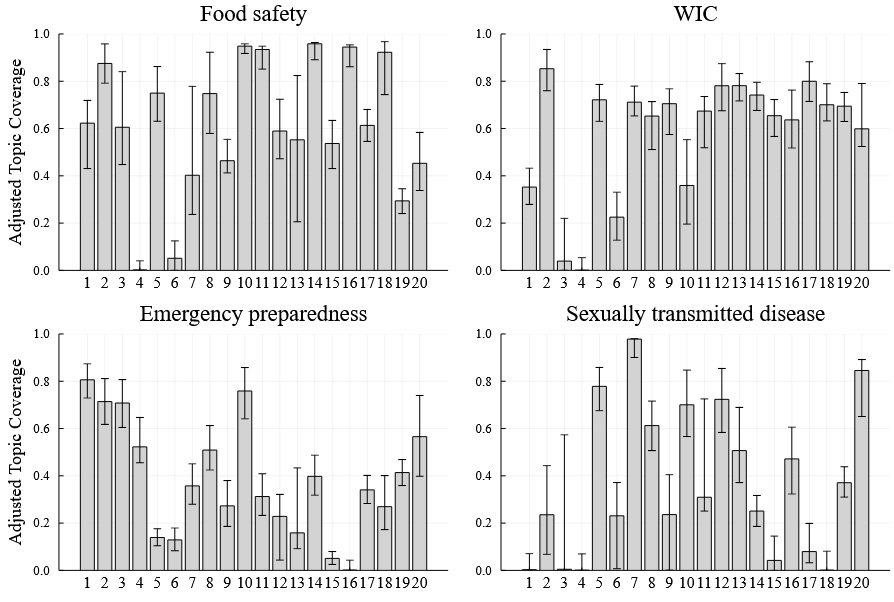}
	\caption{Bar plots of adjusted topic coverage for four global topics from Table \ref{tab:topics}. Bar heights are medians and error bars are 95\% credible intervals.}
	\label{fig:atc20}
\end{figure}

We found that pages in web sites repeat common text, such as geographic names and words, events and news, or contact information. These words have high probability in local topics and local topics account for the largest proportion of web page-topic probability across all web sites. Additionally, the probability of local topics vary between web sites. Thus, we adjust for local topic content on web pages when comparing coverage of (global) health topics. For example, a web page with 20\% probability for its local topic and a 40\% probability for the heart disease topic and a web page with 40\% probability for its local topic and 30\% probability for the heart disease topic should both be viewed as pages 50\% dedicated to the heart disease topic. The adjusted topic coverage (ATC) for topic $k$ in web site $i$ is therefore 
\begin{align*}
	\text{ATC}_{ik} = \underset{j}{\max} ~  \frac{\theta_{ij,k}}{1-\theta_{ij,K+1}}.
\end{align*}
We calculate the adjusted topic coverage for four common health topics, food safety, WIC, emergency preparedness and sexually transmitted disease, using estimates from each of the 1,000 MCMC samples.
Plots of ATC are shown in Figure \ref{fig:atc20}. We use ATC to identify common health topics that may be missing from individual health web sites and in particular investigate web sites where the lower bound of ATC is below 0.05. 

Web sites 4 and 6 have ATC lower bounds below 0.05 for food safety and none of their web pages cover food safety. 
We noted that web sites 7 and 13 have a local topic that shares some high probability words with the food safety topic. However, the ATC for food safety for both web sites are still moderate, between 0.23 and 0.78 in web site 7 and between 0.20 and 0.82. 
For WIC, web site 4 has the lowest ATC and none of its web pages cover WIC.
Web site 3 has ATC lower bound below 0.05 for WIC. The web site mentions WIC in two pages; however, they are not pages dedicated to WIC. One page has 16 frequently asked questions with one related to WIC and another page is an overview of the health department and mentions WIC among other programs and services.
Web site 16 has the lowest ATC for emergency preparedness and, upon inspection, none of its 23 web pages covered emergency preparedness. Web site 15 contains a resource page with multiple sections with one section directing the reader to emergency preparedness web sites outside of web site 15. 

For sexually transmitted diseases (STDs), web sites 1, 3, 4, 15, and 18 have ATC less than 0.05. 
Web sites 3, 4, and 18 did not have web pages covering STDs.
Web site 1 did contain a health information web page with fourteen different drop down menus, each for a different topic. Among the fourteen was an ``STD \& HPV Resource List'' menu.
Web site 15 has a web page listing nine clinical services of which one is a screening and tests service. Under the screening and tests service are 5 tests provided of which one is for STDs and one is for HIV/AIDS screening.
Web sites 6, 9, and 17 additionally have ATC lower bounds below 0.05.
Web site 6 has a page that lists eighteen services that their women's health clinic offers of which one is testing for STDs.
Web site 9 has a page that gives an overview of their reproductive health and wellness clinic and lists services offered. One of the services is testing and treating STDs.
Web site 17 has a page of thirteen frequently asked questions of which one is directly related to STDs. However, testing for STDs is mentioned two additional times as part of larger answers to questions about services offered. This explains why ATC and the ATC lower bound for STDs in web site 17 is the highest of these eight web sites. 

All web sites with ATC lower bound less than 0.05 did not cover the corresponding topic, only linked to an outside resource, or contained a larger page that briefly mentions the topic. ATC looks at a web page's probability of a given topic relative to the cumulative probability of all global topics. Under this metric, a web site with a web page covering several global topics may be considered having low coverage.

\section{Discussion}

\label{discussion}

We introduced and defined local topics as topics that are unique to one web site or group of web pages. Local topics may be common in a nested document collection and we show that in our dataset nearly all local topics included geographical names among their most probable words. We conclude that local topics exist and have high topic probabilities in our dataset. 
We proposed two extensions HA and LT as well as their combination to accommodate the locality and inference in models with nested documents and local topics. 

Adding either or both extensions improves cross-validated log likelihood compared to LDA, and HA-LDA performs better than LT-LDA for larger numbers $K$ of global topics. Combining both extensions, HALT-LDA has a higher peak log likelihood than HA-LDA. However, the peaks are similar between the two and we do not conclude that one outperforms the other in log likelihood. Instead, these two models perform similarly and are both better than LDA or LT-LDA. 
A more notable difference is that HALT-LDA performs well at a smaller number of global topics $K$.
As computation time is largely dependent on the number of topics each word may be drawn from, it is advantageous to use HALT-LDA because it uses smaller $K$ to reach similar performance as HA-LDA.

The key benefit of explicitly modeling local topics is that inference and interpretation are much easier. The model directly identifies local topics and we can infer what proportion of a web page is composed of its local topic. This proportion varies across web sites and web pages. Thus, when comparing coverage of global topics across web sites we should adjust for the probability of local topics.
We compared adjusted topic coverage (ATC) of common health topics across web sites and identified web sites that did not cover food safety, WIC, emergency preparedness, and sexually transmitted disease.

Our goal in modeling nested documents is to study global topics and make comparisons about their distributions within groups of documents. Models should accommodate strong localizations of topics and the addition of local topics and a hierarchical asymmetric prior are useful. However, it may be difficult to determine a priori the number of local topics to introduce. We assumed a single local topic for each web site, which is reasonable for a set of web sites each dedicated to public health in a specific location. However, we noted that 5 web sites in our dataset appear to have two local topics.
We study 5 scenarios in which simulated web sites have none, one, or two local topics in the section A.1 of the web appendix. When local topics are modeled when they do not exist the probability of that local topic is typically small and further, HALT-LDA identifies a local topic that gives high probability to words that occur more often in the local topic's corresponding web site and do not occur as often in the other web sites. When two local topics exist, HALT-LDA almost always merges the two topics into a single local topic. However, this is when the number of global topics $K$ in HALT-LDA matches the number of global topics used to generated the data. When a larger $K$ is set we expect the merged local topic to split as shown in our analysis of 20 web sites with $K = 60$ global toics.

The intervals of conditional posterior means for the highest probability words in topics essentially check for label switching.
Word probabilities for the same word in different common global topics were distinct; if switching were occurring, the 95\% intervals for the word would overlap in the two topics. Thus, the 95\% intervals of the conditional posterior means would be large. The word probabilities shown in Figure \ref{fig:health20} did not fluctuate much which would suggest there was no label switching.
For example, if Food safety and WIC had label-switched, then the 95\% intervals for ``food" would extend from 0.03 to 0.12 in both topics and similarly ``wic" would extend from less than 0.01 to 0.10 in both topics.
\\

\appendix
\section*{Supplemental Materials} 

\subsection*{Web Appendix}
\noindent Web appendix file that includes our simulation study, sensitivity analysis, and brief overview of the left-to-right algorithm applied to HALT-LDA.

\end{document}


\maketitle
\section*{Appendix}
\subsection{Simulation Study of HALT-LDA and Data Sets with Zero, One, or Two Local Topics}
\label{simulation}
We investigate how the HALT-LDA model with one local topic for each web site models simulated data. We are particularly interested in inferences about local topic probabilities either when local topics are not present or when more than one local topic is present in a web site.
All simulated datasets are created with $K = 50$ global topics, $V = 1000$ unique words, $M = 10$ web sites, $M_i = 50$ web pages for all $i = 1,\ldots,10$, and $N_{ij} = 100$ words for each web page $j = 1,\ldots,50$. We study 5 scenarios: (1) no local topics, (2) 5 web sites with one local topic each and no local topics for the remaining web sites, (3) 10 web sites with one local topic each, (4) 5 web sites with one local topic each and 5 with two local topics each, and (5) 10 web sites with two local topics each. 

For web sites with one local topic the non-standardized web site average local topic probabilities $\mu_{i,K+1}$ are generated from a $\text{Normal}(0.25, 0.05^2)$ truncated at 0 and 1. 
Web page topic probabilities $\theta_{ijk}$ are generated by first sampling an unstandardized local topic probability from $\text{Normal}(\mu_{i,K+1}, 0.05^2)$ and unstandardized global topic probabilities from $\text{Dirichlet}(\{0.04,\ldots,0.04\})$ then standardizing such that $\sum_{k=1}^{51}\theta_{ijk} = 1$. We chose an unstandardized mean of 0.25 as 0.2 (0.25/1.25) is a reasonable average local topic probability.
For web sites with two local topics, we generate $\mu_{i,K+1}$ and $\mu_{i,K+2}$ from $\text{Normal}(0.15, 0.05^2)$ and $\text{Normal}(0.1, 0.05^2)$ respectively. Web page topic probabilities are generated similarly with unstandardized local topic probabilities sampled from $\text{Normal}(\mu_{i,K+1},$ $0.05^2)$ and $\text{Normal}(\mu_{i,K+2},$ $0.05^2)$.
Topic-word probabilities are generated from a $\text{Dirichlet}(\{0.01,\ldots,0.01\})$ to get an average highest order statistic around 0.20 and about 20 words with probability greater than 0.01 in each topic. Topic and word indices are sampled from Multinomial distributions given web page topic probabilities and topic-word probabilities.
We generate 100 datasets for each of the 5 variations for a total of 500 simulated datasets.

Web site average local topic probabilities are 
$$\frac{1}{50}\sum_{j=1}^{50}\theta_{ij,K+1}$$
for web sites with one local topic or 
$$\frac{1}{50}\sum_{j=1}^{50}(\theta_{ij,K+1} + \theta_{ij,K+2})$$ 
for web sites with two local topics. Estimates of web site average local topic probability are  $$\frac{1}{Q}\sum_{q=1}^{Q}\frac{1}{50}\sum_{j=1}^{50}\bar{\theta}^{(q)}_{ij,K+1}$$
or
$$\frac{1}{Q}\sum_{q=1}^{Q}\frac{1}{50}\sum_{j=1}^{50}(\bar{\theta}^{(q)}_{ij,K+1} + \bar{\theta}^{(q)}_{ij,K+2})$$
where $Q = 500$ MCMC iterations. There are 10 estimates for web site average local topic probability for each of the 500 simulated datasets.

\begin{figure}
	\centering
	\includegraphics[scale = .7]{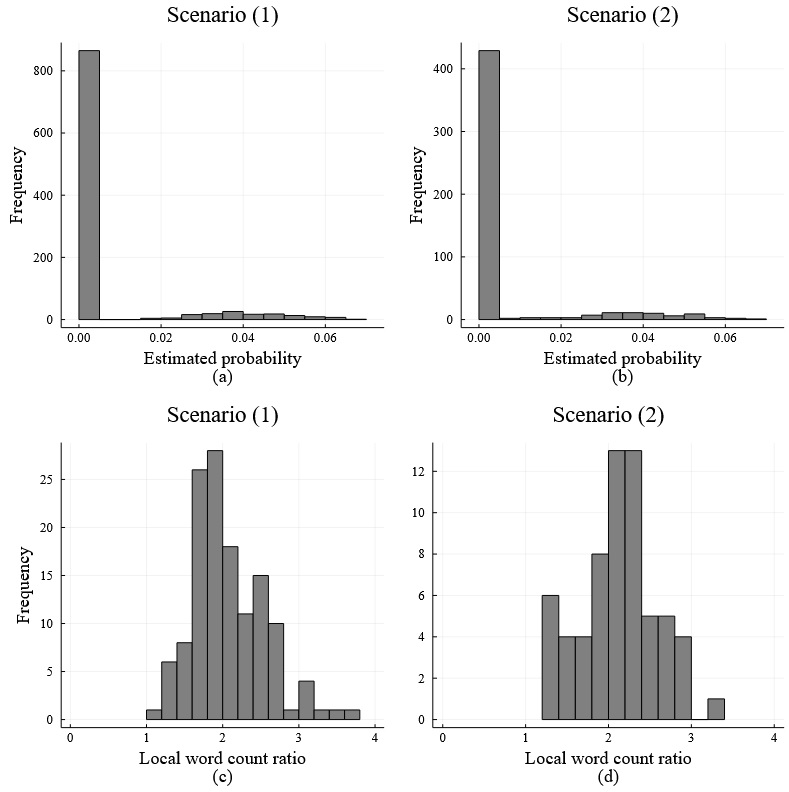}
	\caption{(a) Histogram of estimated web site average local topic probabilities for web sites with no local topic in scenario (1), 10$\times$100 estimates are plotted.
		(b) Histogram of estimated web site average local topic probabilities for web sites with no local topic in scenario (2), 5$\times$100 estimates are plotted.
		(c) and (d) Histograms of local word count ratios of the highest probability words in extraneous local topics in scenario (1) and (2) respectively.}
	\label{fig:wordratio}
\end{figure}

The first row of A.Figure \ref{fig:wordratio} shows histograms of estimated web site average local topic probabilities when local topics are not present. The true web site average local topic probability is 0 and approximately 86\% of all estimates are less than 0.005. When local topics are not present, HALT-LDA typically estimates web site average local topic probability near 0. When estimates are greater than 0.005 they typically range between 0.01 and 0.07, much lower than the 0.20 average. Extraneous local topics with probability estimates greater than 0.02 are further investigated. The 3 highest probability words in each of the extraneous local topics are counted in their corresponding web sites and counted in other web sites then averaged among the other web sites.

Define word count ratio as the ratio of corresponding web site count to other web site count.
Ratios larger than 1 indicate the extraneous local topic has high probability words found more often in its own web site than in other web sites. All local word count ratios are greater than 1. Thus, when local topics do not exist but are modeled, HALT-LDA identifies a topic as local that places high probability on words found more often in the given web site than in other web sites. 

\begin{figure}
	\centering
	\includegraphics[scale = .7]{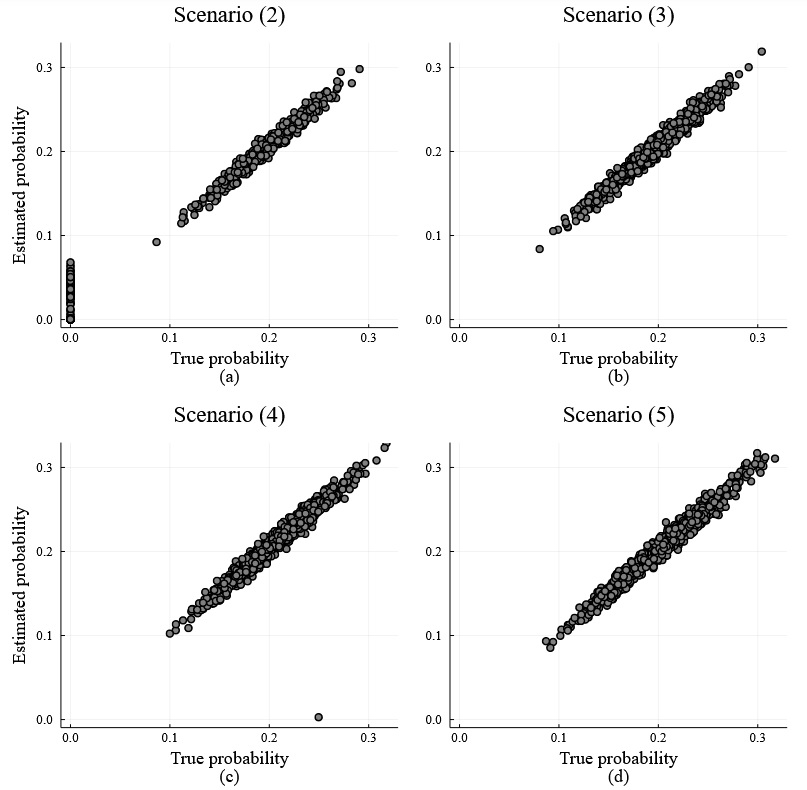}
	\caption{Scatterplots of 1000 estimated web site average local topic probabilities vs true web site average local topic probabilities. (a) Scenario (2) where 5 web sites have one local topic each and 5 web sites have no local topics. 
		(b) Scenario (3) where all 10 web sites have one local topic each.
		(c) Scenario (4) where 5 web sites have two local topics each and 5 web sites have one local topic each.
		(d) Scenario (5) where all 10 web sites have two local topics each.}
	\label{fig:sim_esttrue}
\end{figure}

A.Figure \ref{fig:sim_esttrue} plots estimated web site average local topic probability against true web site average local topic probability for scenarios (2) to (5) where local topics exist. The estimated web site average local topic probabilities are close to the true web site average local topic probability. The bottom two figures indicate that when HALT-LDA models web sites with two local topics, it can merge local topics to a single local topic, when the number of global topics modeled is limited to the true number of global topics. 
We expect some merged local topics to split into the two local topics with one local topic modeled as a global topic if we were to allow more than 50 global topics.

Among the results of all 400 simulated datasets where local topics do exist, only one web site of the 3500 web sites with local topics shows HALT-LDA incorrectly modeling no local topics when there was indeed a local topic present. The point near (0.25,0) A.Figure \ref{fig:sim_esttrue} (c) is for that local topic. The local topic for that web site was instead identified as a global topic.
Inspection of highest probability words of local topics and global topics, as shown in Section 5 of the main text, is recommended to determine that local topics are indeed correctly identified.

\subsection{Sensitivity Analysis of HALT-LDA Conclusions to Prior Specifications}
We study the effects of changing hyperparameters $c_\beta$, $c_\gamma$, and $a_\alpha$ on global topic-word distributions $\phi_{k,v}$. The hyperparameters used in the main results are $c_\beta = c_\gamma = 0.05$ and $a_\alpha = 1$. We consider four sensitivity analysis scenarios doubling or halving these parameters. The first sensitivity analysis model (SA1) sets $c_\beta = c_\gamma = 0.025$ and $a_\alpha = 0.5$; the second sensitivity analysis model (SA2) sets $c_\beta = c_\gamma = 0.025$ and $a_\alpha = 2$; the third sensitivity analysis model (SA3) sets $c_\beta = c_\gamma = 0.1$ and $a_\alpha = 0.5$; the fourth sensitivity analysis model (SA4) sets $c_\beta = c_\gamma = 0.1$ and $a_\alpha = 2$. 
The overall prevalence of local topics in all 5 settings range between 17\% and 19\% with the main results at 18\%, SA1 at 18\%, SA2 at 19\%, SA3 at 17\%, and SA4 at 19\%. 
We match global topics in SA1, SA2, SA3, SA4 to the health topics in Table 2 of the main text with the rank based method using the top 10 words.
A.Table \ref{tab:senstopics} shows the 3 highest probability words for the nine health topics and for the matched topics in the four sensitivity analysis models and the prevalence of the topic.
Generally, the more prevalent topics are similar across all 5 settings. The breast cancer topic differs the most in the 3 most probable words between the main results and the four sensitivity model results. However, when looking at the 10 most probable words from table 2 in the main text, the breast cancer topic from the main results includes all words \textit{screen, cancer, breast, women} found in the 3 most probable words in SA1, SA2, SA3, and SA4. 
The topics modeled by HALT-LDA are fairly robust to changes in hyperparameters $c_\beta$, $c_\gamma$, and $a_\alpha$. 

\begin{table*}
	\caption{Global topics from SA1, SA2, SA3, and SA4 are matched to the health topics shown in Table 2 of the main text using the rank based method with the top 10 words.
		The 3 highest probability words for the nine health topics in each sensitivity analysis and the main analysis are shown with their respective topic prevalences. }
	\label{tab:senstopics}
	\centering
	\setlength{\tabcolsep}{2pt}
	\begin{tabular}{clllll}
		\hline
		& Main & SA1 & SA2 & SA3 & SA4 \\
		\hline
		Disease& \textit{diseas} (3.3\%) & \textit{diseas} (3.2\%) & \textit{diseas} (3.3\%) & \textit{diseas} (3.5\%) & \textit{prevent} (3.7\%)\\	
		prevention& \textit{prevent} & \textit{caus} & \textit{peopl} & \textit{peopl} & \textit{caus}\\
		& \textit{risk} & \textit{peopl} & \textit{caus} & \textit{caus} & \textit{risk} \\
		
		Food& \textit{food} (2.9\%) & \textit{food} (1.8\%) & \textit{food} (1.5\%) & \textit{food} (1.9\%) & \textit{food} (1.7\%)\\	
		safety& \textit{inspect} & \textit{establish} & \textit{inspect} & \textit{establish} & \textit{establish}\\
		& \textit{establish} & \textit{permit} & \textit{establish} & \textit{inspect} & \textit{permit} \\
		
		WIC& \textit{wic} (2.7\%) & \textit{wic} (2.9\%) & \textit{wic} (2.4\%) & \textit{wic} (2.7\%) & \textit{wic} (2.6\%)\\	
		& \textit{breastfeed} & \textit{breastfeed} & \textit{breastfeed} & \textit{breastfeed} & \textit{breastfeed}\\
		& \textit{infant} & \textit{infant} & \textit{infant} & \textit{infant} & \textit{infant} \\
		
		Vaccinations& \textit{immun} (2.0\%) & \textit{vaccin} (2.3\%) & \textit{vaccin} (2.4\%) & \textit{immun} (2.3\%) & \textit{vaccin} (2.2\%)\\	
		& \textit{vaccin} & \textit{immun} & \textit{immun} & \textit{vaccin} & \textit{immun}\\
		& \textit{adult} & \textit{adult} & \textit{adult} & \textit{adult} & \textit{adult} \\
		
		Breast& \textit{test} (1.9\%) & \textit{screen} (0.9\%) & \textit{cancer} (1.0\%) & \textit{screen} (1.2\%) & \textit{cancer} (1.1\%)\\	
		cancer& \textit{women} & \textit{cancer} & \textit{screen} & \textit{women} & \textit{women}\\
		& \textit{clinic} & \textit{women} & \textit{breast} & \textit{cancer} & \textit{screen} \\
		Emergency& \textit{emerg} (1.8\%) & \textit{emerg} (1.8\%) & \textit{emerg} (2.1\%) & \textit{emerg} (1.9\%) & \textit{emerg} (2.2\%)\\	
		preparedness& \textit{prepar} & \textit{prepar} & \textit{prepar} & \textit{prepar} & \textit{prepar}\\
		& \textit{disast} & \textit{disast} & \textit{disast} & \textit{disast} & \textit{disast} \\
		
		Hospital& \textit{care} (1.7\%) & \textit{care} (1.8\%) & \textit{care} (1.8\%) & \textit{care} (1.7\%) & \textit{care} (2.0\%)\\	
		care& \textit{patient} & \textit{inform} & \textit{inform} & \textit{priva} & \textit{provid}\\
		& \textit{provid} & \textit{priva} & \textit{priva} & \textit{medic} & \textit{patient} \\
		
		Sexually& \textit{test} (1.5\%) & \textit{test} (1.5\%) & \textit{test} (0.7\%) & \textit{test} (0.8\%) & \textit{test} (1.3\%)\\	
		transmitted& \textit{std} & \textit{std} & \textit{std} & \textit{std} & \textit{std}\\
		disease& \textit{clinic} & \textit{hiv} & \textit{hiv} & \textit{hiv} & \textit{hiv} \\
		
		Family& \textit{child} (1.4\%) & \textit{child} (1.7\%) & \textit{child} (1.4\%) & \textit{child} (1.5\%) & \textit{child} (1.3\%)\\	
		program& \textit{children} & \textit{program} & \textit{famili} & \textit{famili} & \textit{famili}\\
		& \textit{famili} & \textit{famili} & \textit{children} & \textit{program} & \textit{parent} \\
		\hline
	\end{tabular}	
\end{table*}

\subsection{Left-to-Right Algorithm}
We use the left-to-right algorithm \citep{Wallach2009} and describe how we adapted it to our HALT-LDA model. 
To evaluate our models with the left-to-right algorithm we split pages randomly from each web site into 80\% for MCMC sampling and the remaining 20\% of each web site for evaluation.
For each sample $q$ we calculate 
conditional posterior means $\bar{\phi}_{k,v}^{(q)}$ and $\bar{\psi}_{i,v}^{(q)}$ from the counts in sample $q$ and save the sampled $c_\alpha^{(q)}$ and $\alpha_i^{(q)}$. We calculate an estimate for scale parameter $c_\alpha$, probability vectors $\alpha_i$ and topic-word probabilities $\phi_{k,v}$ and $\psi_{i,v}$ by averaging over 500 MCMC samples after a burn-in of 1500 samples.

The left-to-right algorithm approximates the probability of a held-out document $W_{ij}$ given topic-word probabilities $\phi$ and $\psi_i$ and Dirichlet parameters $c_\alpha$ and $\alpha_i$ or $P(W_{ij} | \phi,\psi_i,c_\alpha,\alpha_i)$.
We provide pseudocode for a single held-out document but it can be extended to multiple held-out documents by adding outer loops over held-out web pages of each web site. The number of particles $R$ is set to $4\times2000/N_{ij}$ as suggested by \citet{Wallach2009}, where $N_{ij}$ is the number of words in web page $j$ in web site $i$.
\begin{enumerate}
	\item initialize $ll = 0$
	\item for h-th word $w_{ijh}$ in held-out document $W_{ij}$ do
	\item ~~~~ initialize $p_{ijh}$ = 0
	\item ~~~~ for each particle $r = 1,\ldots,R$ do
	\item ~~~~~~~~ for $h' < h$ do
	\item ~~~~~~~~~~~~ sample $z_{ijh'}^{(r)}$ from $\text{multinomial}$ on 1:K+1 where \\ 
	$$P(z_{ijh'}^{(r)} = k| w_{ijh'}, \phi, \psi_i, c_\alpha, \alpha, \{z_{h''}^{(r)}\}_{h'' \neq h', h'' < h}) 
	\propto (m_{ij,k}'^{(r)-}+ \alpha_i)
	\phi_{k,w_{ijh}}^{1_{k\leq K}}\psi_{i,w_{ijh}}^{1_{k = K + 1}},$$
	  \indent~~~~~~~~~~~~ for $k = 1,\ldots, K+1$ where $m_{ij,k}'^{(r)-}$ is the count of words from topic \\ \indent~~~~~~~~~~~~ $k$ for all $h'' < h$ where $h'' \neq h'$.
	\item ~~~~~~~~ end for
	\item ~~~~~~~~ $p_{ijh} = p_{ijh} + \sum_{k = 1}^{K+1}P(w_{ijh}, z_{ijh}^{(r)} = k| \phi, \psi_i, c_\alpha, \alpha)$
	\item ~~~~~~~~ sample $z_{ijh}^{(r)}$ from $\text{multinomial}_{1:K+1}$ where\\
	 $$P(z_{ijh}^{(r)} = k | w_{ijh}, \phi, \psi_i, c_\alpha, \alpha, \{z_h' \text{ where } h' < h \})\propto (m_{ij,k}^{(r)-}+ \alpha_i)
	 \phi_{k,w_{ijh}}^{1_{k\leq K}}\psi_{i,w_{ijh}}^{1_{k = K + 1}},
	 $$
	 \indent~~~~~~~~ for $k = 1,\ldots, K+1$  where $m_{ij,k}^{(r)-}$ is the count of words from topic $k$ \\ \indent~~~~~~~~ for all $h' < h$.
	\item ~~~~ end for
	\item ~~~~ $p_{ijh} = p_{ijh}/R$
	\item ~~~~ $ll = ll + \log(p_{ijh})$
	\item end for
	\item $\log(P(W_{ij} | \phi,\psi_i,c_\alpha,\alpha_i)) \approx ll$
\end{enumerate}

\bibliographystyle{apalike}